\documentclass[aps,print,showpacs,a4paper,10pt,twocolumn]{revtex4}
\usepackage{amssymb}
\usepackage{amsmath}
\usepackage{graphicx}

\input{tcilatex}
\begin{document}

\title{Domain Wall Solutions of Spinor Bose-Einstein Condensates in an
Optical Lattice}
\author{Zai-Dong Li$^{1,2}$, Qiu-Yan Li$^{1,2}$, Peng-Bin He$^{3}$, J.-Q.
Liang$^{4}$, W. M. Liu$^{5}$, and Guangsheng Fu$^{2}$}
\affiliation{$^{1}$Department of Applied Physics, Hebei University of Technology, Tianjin
300401, China\\
$^{2}$School of Information Engineer, Hebei University of Technology,
Tianjin, 300401, China\\
$^{3}$College of Physics and Microelectronics Science, Hunan University,
Changsha 410082, China\\
$^{4}$Institute of Theoretical Physics and Department of Physics, Shanxi
University, Taiyuan 030006, China\\
$^{5}$Beijing National Laboratory for Condensed Matter Physics, Institute of
Physics, Chinese Academy of Sciences, Beijing 100080, China }

\begin{abstract}
We studied the static and dynamic domain wall solutions of spinor
Bose-Einstein condensates trapped in an optical lattice. The single and
double domain wall solutions are constructed analytically. Our results show
that the magnetic field and light-induced dipolar interactions play an
important role for both the formation of different domain walls and the
adjusting of domain wall width and velocity. Moreover, these interactions
can drive the motion of domain wall of Bose ferromagnet systems similar to
that driven by the external magnetic field or the spin-polarized current in
fermion ferromagnet.
\end{abstract}

\maketitle

During the past several decades, the dynamics of spatial domain wall have
attracted more attention in ferromagnetism \cite{Mohn,Volkov}. The
conventional ferromagnet is usually composed of fermions which contribute
the main site-to-site exchange interaction owing to the direct Coulomb
interaction among electrons and Pauli exclusion principle. This exchange
interaction causes spin wave unstable, and its developing instability brings
about the appearance of spatial domain walls or magnetic soliton \cite{Kose}%
. The contribution of the magnetic dipole-dipole interaction to domain wall
formation is usually neglected in practice because it is typically several
orders of magnitude weaker than the exchange coupling in fermion ferromagnet.

However, the description of magnetism composed of Bosons are not well
explored. Fortunately, a typical Bose system was realized in cold spinor $%
^{87}$Rb gases \cite{Stenger,Anderson} and $^{23}$Na gases \cite{Stamper}
which provided a totally new environment to understand magnetism
comprehensively. The ferromagnetic spinor Bose gases have attracted numerous
theoretical and experimental interests \cite%
{Ho,Miesner,Pu,Zhang,Strecker,Kasa,Sarlo,Wxz,Sadler,Gu}, and some general
properties as in conventional fermion ferromagnets have been observed, such
as spontaneous symmetry-broken ground state \cite{Ho}, spin domains and
textures \cite{Miesner}, and normal spin-wave excitations spectrum \cite{Pu}%
. Especially, the spontaneous symmetry breaking in $^{87}$Rb spinor
condensates \cite{Sadler} clearly shows ferromagnetic domains and domain
walls in a Bose ferromagnet. Recently, the ferromagnetism in a Fermi gas is
also of interest \cite{fermi} in the field of ultracold atoms.

Spinor Bose-Einstein condensates (BECs) trapped in an optical potential have
internal degrees of freedom due to the hyperfine spin of the atoms which
liberate a rich variety of phenomena. When the potential valley is so deep
that the individual sites are mutually independent, spinor BECs at each
lattice site behave like spin magnets and can interact with each other
through both the magnetic and the light-induced dipole-dipole interactions
which are no longer negligible due to the large number of atoms at each
lattice site. These site-to-site dipolar interactions can cause the
ferromagnetic phase transition \cite{Pu} leading to a \textquotedblleft
macroscopic\textquotedblright \ magnetization of the condensate array and
the spin-wave like excitation \cite{Pu,Zhang} and magnetic soliton \cite%
{lizd,Lu} analogous to the ferromagnetism in solid-state physics, but occur
with bosons instead of fermions. Also, spinor BECs in an optical lattice is
a typical physical realization of Salerno model \cite{Saler}.

In this paper, we explore the domain wall solutions of spinor BECs trapped
in an optical lattice and the roles of the magnetic and light-induced
dipole-dipole interactions for the type of domain wall solutions and domain
wall width and velocity.

We consider $F=1$ spinor condensates trapped in a one-dimensional optical
lattice formed by two-polarized laser beams counter propagating along the $y$%
-axis and the two laser beams are detuned far from atomic resonance. Under
the tight-binding approximation, the Hamiltonian takes the form \cite%
{Ho,Miesner,Pu,Zhang}%
\begin{eqnarray}
\hat{H} &=&\sum_{n}[\lambda _{a}^{\prime }\mathbf{\hat{S}}_{n}^{2}-\gamma
_{B}\hat{S}_{n}\cdot \mathbf{B}-\sum_{l\neq n}J_{nl}^{\text{iso}}\mathbf{%
\hat{S}}_{n}\cdot \mathbf{\hat{S}}_{l}  \notag \\
&&-\sum_{l\neq n}J_{nl}^{\text{tran}}\left( \hat{S}_{n}^{+}\hat{S}_{l}^{-}+%
\hat{S}_{n}^{-}\hat{S}_{l}^{+}\right) ],  \label{ham2}
\end{eqnarray}%
where $\mathbf{\hat{S}}_{n}$ is the $n$th collective spin operator, defined
as $\mathbf{\hat{S}}_{n}=\hat{a}_{\alpha }^{\dag }(n)\mathbf{F}_{\alpha
\upsilon }\hat{a}_{\upsilon }(n)$, with $\hat{a}$ being annihilation
operator and $\mathbf{F}$ being the vector operator for the hyperfine spin
of an atom. The first term\ in\textbf{\ }the Hamiltonian results from the
spin-dependent interatomic collisions at a given lattice site, with $\lambda
_{a}^{\prime }=(1/2)\lambda _{a}\int d^{3}r\left \vert \phi _{n}(\mathbf{r}%
)\right \vert ^{4}$, where $\lambda _{a}$ is proportional to the difference
between the s-wave scattering lengths in the triplet and singlet channels 
\cite{Ho}, and $\phi _{n}(\mathbf{r})$ is the ground-state wave function for 
$n$th site. The direction of the magnetic field $\mathbf{B}$ is along $z$%
-axis and $\gamma _{B}=\mu _{B}g_{F}$ is the gyromagnetic ratio, with $\mu
_{B}$ being the Bohr magneton and $g_{F}$ the Land\'{e} factor. The last two
terms describe the site-to-site spin coupling induced by both the static
magnetic and the light-induced dipole-dipole interactions \cite{Zhang}, with
the form $J_{nl}^{\text{iso}}=J_{nl}^{\prime }+J_{nl}^{\prime \prime }$,
where $J_{nl}^{\prime }=\mu _{0}\gamma _{B}^{2}/(16\pi \hbar ^{2})\tiint d%
\mathbf{r}d\mathbf{r}^{\prime }\left \vert \phi _{n}(\mathbf{r})\right \vert
^{2}\left \vert \phi _{l}(\mathbf{r-r}^{\prime })\right \vert
^{2}/\left
\vert \mathbf{r}^{\prime }\right \vert ^{3}$, $J_{nl}^{\prime
\prime }=-3\mu _{0}\gamma _{B}^{2}/(16\pi \hbar ^{2})\tiint d\mathbf{r}d%
\mathbf{r}^{\prime }\left \vert \phi _{n}(\mathbf{r})\right \vert
^{2}\left
\vert \phi _{l}(\mathbf{r-r}^{\prime })\right \vert ^{2}y^{\prime
2}/\left
\vert \mathbf{r}^{\prime }\right \vert ^{5}$, $J_{nl}^{\text{tran}%
}=\gamma U_{0}/(24\Delta \hbar ^{2}k_{L}^{3})\tiint d\mathbf{r}d\mathbf{r}%
^{\prime }f_{c}\left( \mathbf{r}^{\prime }\right) e^{(-r_{\bot
}^{2}-\left
\vert \mathbf{r}_{\bot }-\mathbf{r}_{\bot }^{\prime
}\right
\vert ^{2})/W_{L}^{2}}$ $\cos \left( k_{L}y-k_{L}y^{\prime }\right)
\cos (k_{L}y)\mathbf{e}_{+1}\cdot \mathbf{W}(\mathbf{r}^{\prime })\cdot 
\mathbf{e}_{-1}\left \vert \phi _{l}(\mathbf{r-r}^{\prime })\right \vert
^{2} $ $\left
\vert \phi _{n}(\mathbf{r})\right \vert ^{2}+3J_{nl}^{\prime
\prime }/2$. Here the cutoff function $f_{c}\left( \mathbf{r}\right)
=e^{-r^{2}/L_{c}^{2}} $ describes the effective interaction range of the
light-induced dipole-dipole interaction, with $L_{c}$ being the coherence
length associated with different decoherence mechanisms \cite{Java} and $%
\Delta $ being the detuned frequency. The wave number $k_{L}=2\pi /\lambda
_{L}$, the transverse coordinate $r_{\bot }=\sqrt{x^{2}+z^{2}}$, $W_{L}$ is
the width of the lattice laser beams, and $U_{0}$ denotes the depth of
optical lattice potential. The $\mathbf{e}_{\pm 1}$ are unit vectors in the
spherical harmonic basis, and the form of tensor $\mathbf{W}\left( \mathbf{r}%
\right) $ can be found in Ref. \cite{Zhang}. Eq. (\ref{ham2}) shows that the
static magnetic and the light-induced dipole-dipole interaction can lead to
the isotropic spin coupling denoted by $J_{nl}^{\text{iso}}$ and anisotropic
spin coupling in the transverse direction denoted by $J_{nl}^{\text{tran}}$.

From Hamiltonian (\ref{ham2}) we can derive the Heisenberg equation of
motion at $k$th site for the spin, i.e., $i\hbar \partial \hat{S}%
_{k}/\partial t=[\hat{S}_{k},\hat{H}]$. When the optical lattice is
infinitely long and in the ultra-low temperatures for condensation, the
operator can be treated as a classical vector, $\mathbf{\hat{S}}%
_{k}\rightarrow \mathbf{S}\left( y,t\right) $. Here we assume all
nearest-neighbor interactions are same, which is a good approximation in
one-dimensional optical lattice \cite{Konotop}. Then we get the effective
Landau-Lifshitz equation \cite{lizd}%
\begin{equation}
\frac{\partial }{\partial t}\mathbf{S}=\frac{1}{\hbar }\mathbf{S}\times
\lbrack 2Jd_{0}^{2}(\frac{\partial ^{2}\mathbf{S}}{\partial y^{2}}\mathbf{-}%
\frac{4J^{\text{tran}}}{J}S^{z}\mathbf{e}_{z})+\gamma B\mathbf{e}_{z}],
\label{magnet}
\end{equation}%
where $J=2J^{\text{tran}}+J^{\text{iso}}$ and $d_{0}=\lambda _{L}/2$ being
the lattice constant. In a rotating frame around $z$-axis with angular
frequency $\gamma _{B}B/\hbar $\ the spin vector $\mathbf{S}$\ is related
with the original one by the transformation $S_{+}\equiv
S^{x}+iS^{y}=S_{+}^{\prime }e^{-i\gamma _{B}Bt/\hbar }$. Thus Eq. (\ref%
{magnet}) becomes%
\begin{equation}
\frac{\partial }{\partial t}\mathbf{S}=\mathbf{S}\times \left( \frac{%
\partial ^{2}}{\partial y^{2}}\mathbf{S-}\beta S^{z}\mathbf{e}_{z}\right) ,
\label{LL1}
\end{equation}%
where the superscript ($^{\prime }$)\ is omitted for pithiness, and the time 
$t$ and coordinate $y$ is in unit $t_{0}=\hbar /\left( 2J\right) $ and $%
d_{0} $, respectively. The parameter $\beta \equiv 4J^{\text{tran}}/J$ can
be controlled by tuning the lattice laser beams and the shape of the
condensate at each lattice site. Such a character makes the lattice atomic
spin chain an ideal tool to study a diversity of spin-related phenomena.

\textbf{Static domain wall solutions: }We firstly seek for the single domain
wall solution in the form%
\begin{equation}
S^{x}=\frac{\cos \eta _{0}^{\prime \prime }}{\cosh \Theta _{0}},S^{y}=\frac{%
\sin \eta _{0}^{\prime \prime }}{\cosh \Theta _{0}},S^{z}=\tanh \Theta _{0},
\label{kink1a}
\end{equation}%
where $\Theta _{0}=k_{0}y-\omega _{0}t+\eta _{0}^{\prime }$ with $\eta
_{0}^{\prime }$, $\eta _{0}^{\prime \prime }$ being two real parameters, and 
$k_{0}$, $\omega _{0}$ to be determined. Substituting Eq. (\ref{kink1a})
into Eq. (\ref{LL1}) we find that $\omega _{0}=0$ and $k_{0}=\sqrt{-\beta }$%
, 
\begin{figure}[tbp]
\begin{center}
\includegraphics[width=0.7\linewidth]{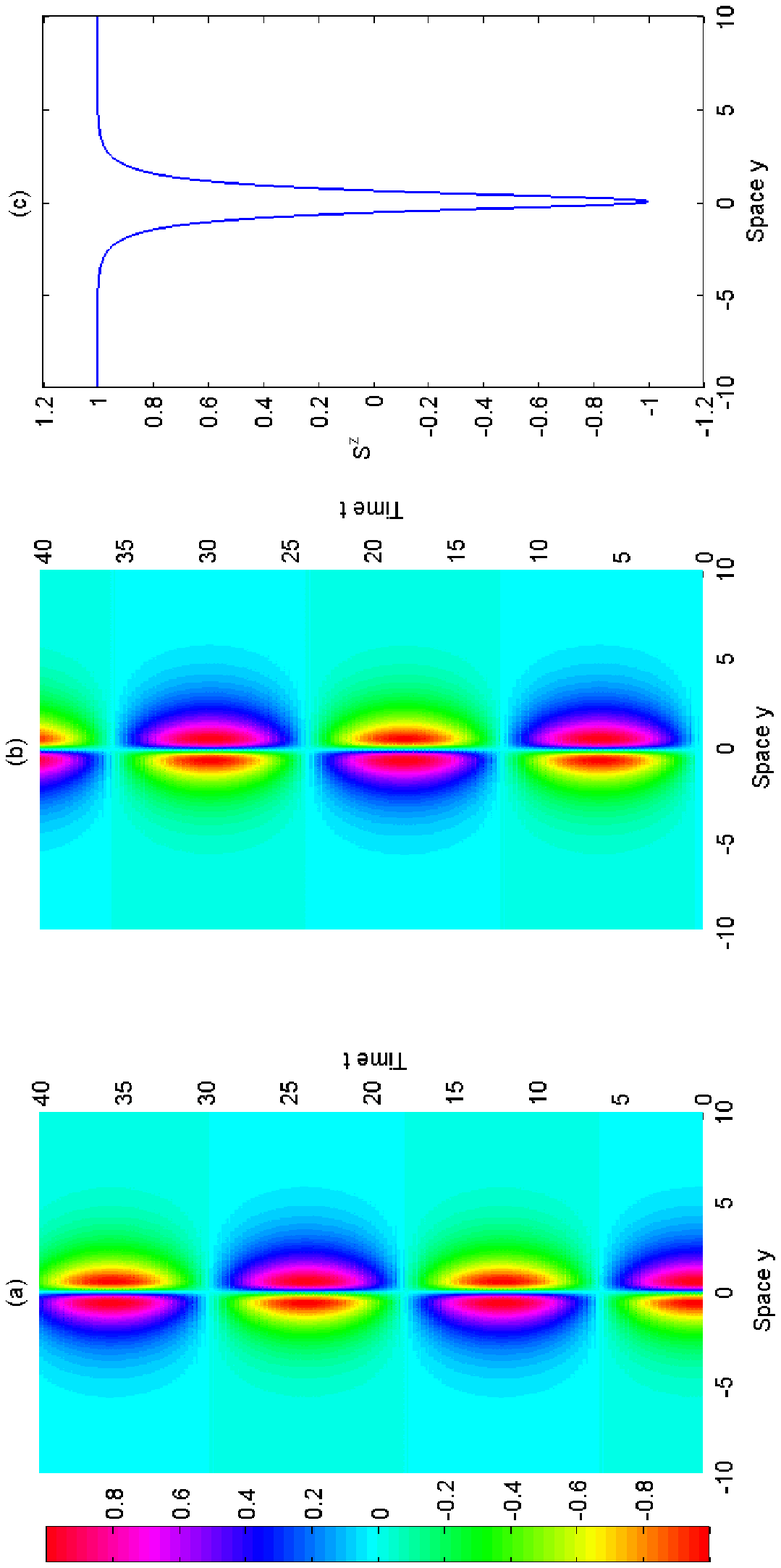}
\end{center}
\caption{(Color online) The illustration of the double domain solutions (a) $%
S^{x}$, (b) $S^{y}$, and (c) $S^{z}$ in Eq. (\protect \ref{DK1a}) with
condition $\protect \beta =-0.2$ and $k_{1}=0.3$. The two components $S^{x}$
and $S^{y}$ possess of the double peak located at $\sinh \Theta _{1}=\pm 1/%
\protect \delta _{1}$.}
\end{figure}
and the solution (\ref{kink1a}) in fact denotes the static domain wall. This
result implies the parameter $\beta <0$ which can be realized by a
blue-detuned lattice $\left( \Delta <0\right) $, where the condensed atoms
are trapped at the standing-wave nodes and the laser intensity is
approximately zero. As a result, the light-induced dipole-dipole
interactions is very small and the spin coupling is governed mainly by the
static magnetic dipole-dipole interaction which implies $\left \vert \beta
\right \vert <<1$. In this case the domain wall width, defined by $1/k_{0}$,
can be adjusted mainly by $J^{\text{iso}}$. Under the condition $J^{\text{%
tran}}/J^{\text{iso}}\approx -0.25$, the domain wall width is about $%
\allowbreak 3d_{0}$ with $d_{0}$ being the lattice constant.

Eq. (\ref{LL1}) has a norm form of Landau-Lifshitz type for a spin chain
with an anisotropy. It can be solved by inverse scattering method where a
couple of Lax equations is introduced for constructing the analytical
solutions. In terms of our earlier results \cite{Licpl} we get two static
double domain wall solutions. The first has the form%
\begin{eqnarray}
S^{x} &=&-2\delta _{1}\left( \cos \Phi _{1}\sinh \Theta _{1}\right) /\Delta
_{1},  \notag \\
S^{y} &=&-2\delta _{1}\left( \sin \Phi _{1}\sinh \Theta _{1}\right) /\Delta
_{1},  \notag \\
S^{z} &=&1-2/\Delta _{1},  \label{DK1a}
\end{eqnarray}%
where $\Delta _{1}=1+\delta _{1}^{2}\sinh ^{2}\Theta _{1}$, $\Theta
_{1}=[\beta /\left( 8k_{1}\right) -2k_{1}]y$, $\Phi _{1}=[2k_{1}+\beta
/\left( 8k_{1}\right) ]^{2}t$, $\delta _{1}=\left( 16k_{1}^{2}+\beta \right)
/\left( 16k_{1}^{2}-\beta \right) $, and $k_{1}$ is a real parameter. The
illustration of domain wall solution in Eq. (\ref{DK1a}) is shown in Fig. 1.
From Fig. 1 and Eq. (\ref{DK1a}) we see that the two components $S^{x}$ and $%
S^{y}$ precess around the component $S^{z}$ with the frequency $\left(
2k_{1}+\beta /\left( 8k_{1}\right) \right) ^{2}$, and the $z$-component of
spin vector $S^{z}$ is conservative. The component $S^{x}$(or $S^{y}$)
possess of the double peak located at $\sinh \Theta _{1}=\pm 1/\delta _{1}$
which varies with time periodically as shown in Fig. 1. The width of double
domain wall is $1/\left( \beta /\left( 8k_{1}\right) -2k_{1}\right) $, which
is inverse proportion to the parameter $\beta $, while the maximum absolute
value of valley for $S^{z}$ is constant. 
\begin{figure}[tbp]
\begin{center}
\includegraphics[width=0.7\linewidth]{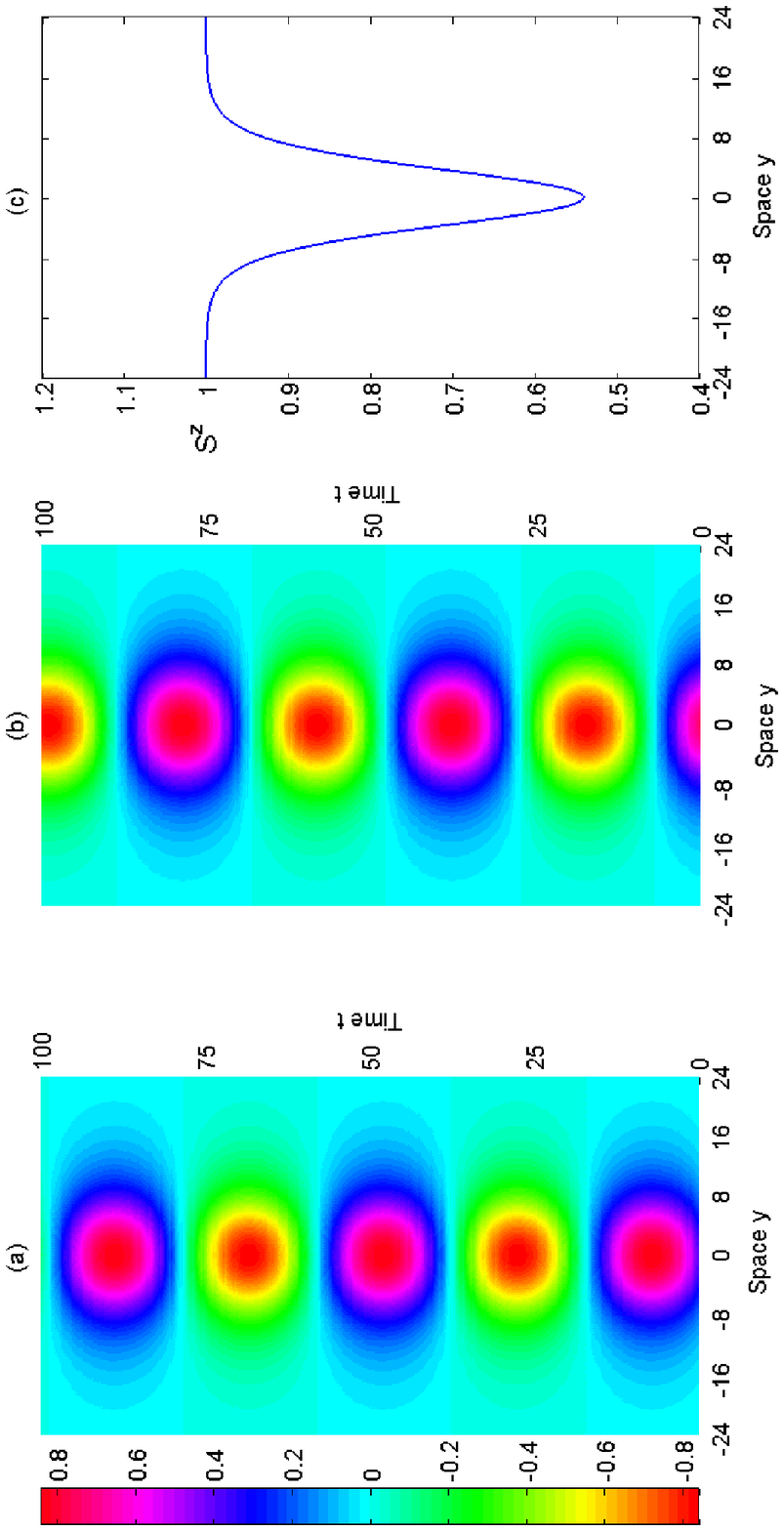}
\end{center}
\caption{(Color online) The illustration of the double domain solutions (a) $%
S^{x}$, (b) $S^{y}$, and (c) $S^{z}$ in Eq. (\protect \ref{DK1b}) with
condition $\protect \beta =-0.2$ and $\protect \varphi =0.5$. The two
components $S^{x}$ and $S^{y}$ possess of the single peak located at $z=0$.}
\end{figure}

The other solution can be written as%
\begin{eqnarray}
S^{x} &=&2[\cot \varphi \sin \left( \varphi -\Phi _{2}\right) \cosh \Theta
_{1}^{\prime }]/\Delta _{1}^{\prime },  \notag \\
S^{y} &=&2[\cot \varphi \cos \left( \varphi -\Phi _{2}\right) \cosh \Theta
_{1}^{\prime }]/\Delta _{1}^{\prime },  \notag \\
S^{z} &=&1-2/\Delta _{1}^{\prime },  \label{DK1b}
\end{eqnarray}%
where $\Delta _{1}^{\prime }=1+\cot ^{2}\varphi \cosh ^{2}\Theta
_{1}^{\prime }$, $\Phi _{2}=\beta \left( \cos ^{2}\varphi \right) t$, $%
\Theta _{1}^{\prime }=-\sqrt{-\beta }\left( \sin \varphi \right) y$ , and $%
\varphi $ is a real parameter. The illustration of domain wall solution in
Eq. (\ref{DK1b}) is shown in Fig. 2. Different from the former one, the two
components $S^{x}$ and $S^{y}$ possess of the single peak located at $y=0$
which also oscillates with time periodically as shown in Fig. 2. The width
of double domain wall is $1/(\sqrt{-\beta }\sin \varphi )$, which is inverse
proportion to $\sqrt{-\beta }$. The maximum absolute value of valley for $%
S^{z}$ is $1-2/\left( 1+\cot ^{2}\varphi \right) $ which is also independent
on the parameter $\beta $. We can rewrite $S^{z}$ in Eqs. (\ref{DK1a}) and (%
\ref{DK1b}) as $S^{z}=((\delta _{1}^{2}+1)\tanh ^{2}\Theta _{1}-1)/((\delta
_{1}^{2}-1)\tanh ^{2}\left( -\Theta _{1}\right) -1)$ and $(\cot ^{2}\varphi
-1+\tanh ^{2}\Theta _{1}^{\prime })/(\cot ^{2}\varphi +1-\tanh ^{2}\Theta
_{1}^{\prime })$, respectively, which implies the double domain wall
solutions can be written asymptotically as a nonlinear combination of two
single domain wall solution in Eq. (\ref{kink1a}).

For observation of the above domain wall, one of the possible physical
realizations of a gas of dipolar BECs can be provided by electrically
polarized gases of polar molecules or by applying a high dc electric field
to atoms \cite{Marinescu}. In order to induce the dipole moment of the order 
$0.1$ D (Debye) one needs an electric field of the order of $10$ V/cm and
the corresponding s-wave scattering length $10$-$1000$ \AA .

\textbf{Dynamic domain wall solutions: }For red-detuned lattices, the
condensed atoms are trapped at the maxima of the intensity of the standing
wave laser and the spin coupling is dominantly determined by the
light-induced dipole-dipole interaction. In particular, the spin coupling is
anisotropic in this case. By controlling the laser parameters, we may always
make the light-induced dipole-dipole interaction dominate over the static
magnetic dipole-dipole interaction, i.e., $J^{\text{tran}}>>J^{\text{iso}}>0$
and $\beta \sim 2$. 
\begin{figure}[tbp]
\begin{center}
\includegraphics[width=0.7\linewidth]{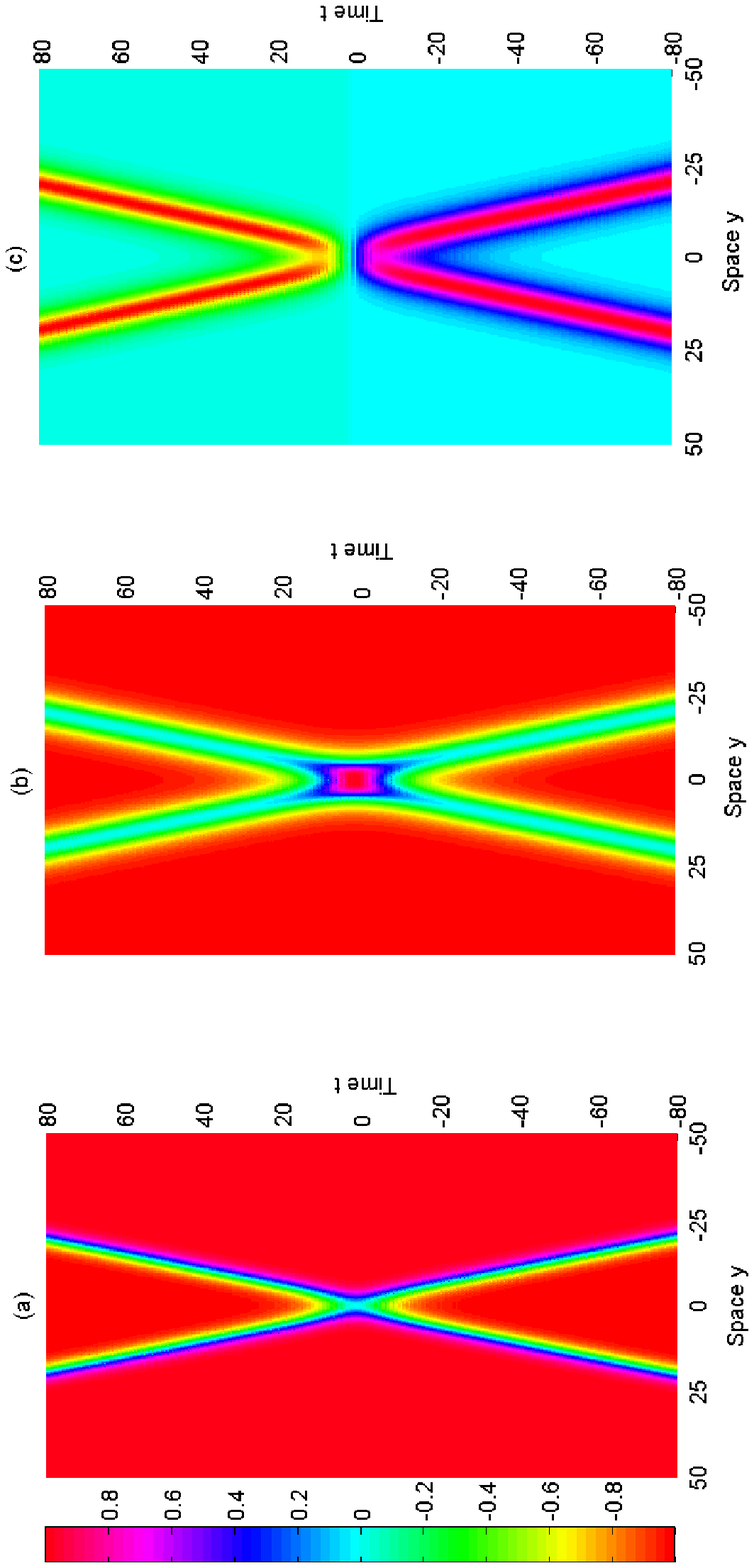}
\end{center}
\caption{(Color online) The scattering process of two moving domain walls
(a) $S^{x}$, (b) $S^{y}$, and (c) $S^{z}$ expressed in Eq. (\protect \ref{DK2}%
), in which two domain walls propagate with the opposite direction and the
same absolute value of domain wall velocity in the condition $\protect \beta %
=1.5$ and $\protect \kappa =\protect \pi /3$.}
\end{figure}
In this case we assume that the single moving domain wall solution of Eq. (%
\ref{LL1}) admits the form%
\begin{equation}
S^{x}=\tanh \Theta _{2},S^{y}=\frac{\cos \eta _{1}^{\prime \prime }}{\cosh
\Theta _{2}},S^{z}=\frac{\sin \eta _{1}^{\prime \prime }}{\cosh \Theta _{2}},
\label{kink1b}
\end{equation}%
where $\Theta _{2}=k_{2}y+\omega _{2}t+\eta _{1}^{\prime }$, with $\eta
_{1}^{\prime }$, $\eta _{1}^{\prime \prime }$ being two real parameters, and 
$k_{2}$, $\omega _{2}$ to be determined. Substituting Eq. (\ref{kink1b})
into Eq. (\ref{LL1}) we obtain $\omega _{2}=\beta /2\sin \left( 2\eta
_{1}^{\prime \prime }\right) $ and $k_{2}=\pm \sqrt{\beta }\sin \eta
_{1}^{\prime \prime }$. The solution (\ref{kink1b}) shows that the dynamics
of domain wall is restricted in ($k_{2},\omega _{2}$)-space, with $\left
\vert k_{2}\right \vert _{\max }=\sqrt{\beta }$ and $\left \vert \omega
_{2\max }\right \vert =\beta /2$. The width of domain wall, defined by $%
1/\left \vert k_{2}\right \vert $, is inverse proportion to the square root
of parameter $\beta $ determined mainly by light-induced dipole-dipole
interaction and the angle $\eta _{1}^{\prime \prime }$ of the three
components of pseudospin vector $\mathbf{S}$. The domain wall velocity,
i.e., $v=\omega _{2}/k_{2}=\pm \sqrt{\beta }\cos \eta _{1}^{\prime \prime }$%
, is dependent on the parameters $\beta $ and $\eta _{1}^{\prime \prime }$.
When $\eta _{1}^{\prime \prime }=n\pi $, $n=0,1,2,...,$ the domain wall
velocity attains its maximum value $v_{\max }=\sqrt{\beta }$. $\allowbreak $%
When $\eta _{1}^{\prime \prime }=\left( n+1/2\right) \pi ,n=0,1,2,...$, the
solution in Eq. (\ref{kink1b}) represents the static domain wall solution of
Eq. (\ref{LL1}) similar the case of $\beta <0$. It is interesting to
estimate the velocity of domain wall. As a example, we consider the $F=1$
electronic ground state of $^{87}$Rb which the Land\'{e} factor $g_{F}=-1/2$
and the gyromagnetic ratio $\gamma _{B}=-\mu _{B}/2$. As from Ref. \cite{Pu}
we estimate $J^{\text{iso}}$ is about $1.\allowbreak 1\times 10^{-36}$J with 
$\lambda _{L}=852$ nm. Asumming that $J^{\text{tran}}=10J^{\text{iso}}$ we
have $\beta \approx 1.\allowbreak 9$ and the maximum domain wall velocity is
about $v=$ $\sqrt{\beta }d_{0}/t_{0}\approx \allowbreak 130$ nm/s. For
chromium atoms, it has a magnetic moment $6\mu _{B}$ and the corresponding
domain wall velocity is about tens of domain wall velocity of $^{87}$Rb with
the same assumption.

The above results show that the domain wall velocity can be controlled by
the parameter $\beta $ resulting from the external magnetic and light
field-induced dipole-dipole interaction and the direction of pseudospin
vector $\mathbf{S}$. In fermion ferromagnet, a magnetic domain wall is a
spatially localized configuration of magnetization, in which the direction
of magnetic moments gradually inverses. When a spin-polarized electric
current flows through a domain wall, the spin-polarization of conduction
electrons can transfer spin momentum to the local magnetization, thereby
applying a spin-transfer torque, which manipulates the motion of domain wall 
\cite{Slon,Katine} similar to that driven by applied an external magnetic
field \cite{Walker}. It is interesting that our domain wall solution in Eq. (%
\ref{kink1b}) shows the same properties as that in fermion ferromagnet,
i.e., for red-detuned lattices the light-induced dipolar interactions can be
seen as the external force to drive the motion of domain wall, and it has
the potential contribution for the research of domain wall motion in the
Bose ferromagnet systems.

The dynamic double domain wall solution can be constructed by employing
Hirota method \cite{Mic} which is an effective straightforward technique to
solve the nonlinear equations. Here we only mention the main idea of Hirota
method briefly. Firstly, it applies a direct transformation to the nonlinear
equation. Then, by means of some skillful bilinear operators the nonlinear
equation can be decoupled into a series of equations. With some reasonable
assumptions the exact solutions can be constructed effectively. Performing
the normal procedure in Ref. \cite{Mic} we get dynamic double domain wall
solution as%
\begin{eqnarray}
S^{x} &=&[e^{\theta }\cosh ^{2}\left( k_{2}^{\prime }y\right) -\sinh
^{2}\left( \omega _{2}^{\prime }t\right) -\cos ^{2}\kappa ]/\Delta _{2}, 
\notag \\
S^{y} &=&2e^{\theta /2}\cos \kappa \cosh \left( k_{2}^{\prime }y\right)
\cosh \left( \omega _{2}^{\prime }t\right) /\Delta _{2},  \notag \\
S^{z} &=&-2e^{\theta /2}\sin \kappa \cosh \left( k_{2}^{\prime }y\right)
\sinh \left( \omega _{2}^{\prime }t\right) /\Delta _{2},  \label{DK2}
\end{eqnarray}%
where $\Delta _{2}=e^{\theta }\cosh ^{2}\left( k_{2}^{\prime }y\right)
+\sinh ^{2}\left( \omega _{2}^{\prime }t\right) +\cos ^{2}\kappa $ with the
real parameter $\kappa $, $k_{2}^{\prime }=\pm \sqrt{\beta }\sin \kappa $, $%
\omega _{2}^{\prime }=\beta /2\sin \left( 2\kappa \right) $, and $\theta
=\ln \left( \cos ^{2}\kappa \right) $. In figure 3 we plot the evolution\ of
double domain wall solution $\mathbf{S}$ in Eq. (\ref{DK2}). From figure 3
we see that the double domain solution in Eq. (\ref{DK2}) presents a general
scattering process of two moving domain walls which propagate with the
opposite direction and the same absolute value of domain wall velocity $%
\sqrt{\beta }\cos \kappa $. Analysis reveals that there is no amplitude
exchange among the components $S^{x}$ for two domain walls. However, there
is a small shift for the center of each domain wall during collision.

In conclusion, we have presented two kinds of domain wall solutions of
dipolar spinor BECs in an optical lattice based on an effective Hamiltonian
of anisotropic pseudospin chain. In real experiment, the magnetic field and
light-induced dipolar interactions are related with the detuned term which
can be controlled easily by the external magnetic and light field. Our
results show that we can control this site-to-site spin coupling of the
condensates at each lattice to form the different domain wall solutions.
Especially, since these magnetic field and light-induced dipolar
interactions are highly controllable, the spinor BECs in an optical lattice
as an exceedingly clean system offer a very useful tool to study spin
dynamics in periodic structures and to understand ferromagnetism
comprehensively.

\textbf{Acknowledgements: }This work was supported by NSF of China under
Grants No. 10874038 and No. 10804028 and the NSF of

Hebei Province of China under Grant No. A2007000006. J.-Q. Liang was
supported by NSF of China under Grant No. 10775091. W. M. Liu was supported
by NSFC under Grants No. 60525417, No. 10740420252, and No. 10874235 and by
NKBRSFC under Grants No. 2006CB921400 and No. 2009CB930704.

\end{document}